\documentclass[showpacs,aps,graphicx,twocolumn]{revtex4}
\usepackage{graphicx}

\begin{document}


\title{Transitionless intra-cavity quantum state transfer in optomechanical systems}

\author{Hao Zhang, Xue-Ke Song, Qing Ai, Mei Zhang, and Fu-Guo Deng\footnote{Corresponding author: fgdeng@bnu.edu.cn}}

\address{Department of Physics, Applied Optics Beijing Area Major Laboratory,
Beijing Normal University, Beijing 100875, China}

\date{\today }

\begin{abstract}
Quantum state transfer between cavities is crucial for quantum information processing and quantum computation in optomechanical systems. Here, we present the first scheme for the  transitionless intra-cavity quantum state transfer based on transitionless quantum driving (TQD) algorithm in optomechanical systems. We also present a physically feasible system for the TQD process based on largely detuned optomechanical cavity. With the Gaussian time-dependence coupling strengths, our scheme can achieve the perfect quantum state transfer with no undesired transition and reduce the dependence of accurately controlling evolution time and interval of coupling strengths. Our computational results show that the TQD process can be accomplished with no need of the mechanical oscillator in its ground state and is also robust to the mechanical dissipation.
\end{abstract}

 \maketitle


\section{Introduction}\label{sec1}

Quantum state transfer is extremely important in quantum information processing and quantum computation \cite{MANielsen2000}. Near perfect quantum state transfer could be realized via adiabatic process, such as rapid adiabatic passage \cite{NVVitanovARPC2001} and the stimulated Raman adiabatic passage technique \cite{KBergmannRMP1998} for two-level and three-level quantum systems, respectively. According to the adiabatic theorem \cite{MBorn1928}, if the state of a quantum system remains non-degenerate and starts in one of the instantaneous eigenstates, it will evolve along this initial state all the time. This process requires the evolution slow enough and it is described by a small parameter $\varepsilon$. Therefore the transition amplitude is very small with the order $exp(-constant/\varepsilon)$ \cite{LLandau1932,CZener1932,EMajorana1932,JPDavis1976,JTHwang1977}. This transition will lead to a certain probability of qubit error in quantum state engineering. In order to achieve the same final state with transitionless process, there are two potentially equivalent approaches to achieve this goal, i.e., Lewis-Riesenfeld invariant-based inverse engineering \cite{HRLewisJMP1969,JGMugaJPB2009,XChenPRA2011,ETorronteguiAAMOP2013,YHChenPRA2014} and transitionless quantum driving (TQD) \cite{MDemirplakJPCA2003,MDemirplakJPCB2005,MDemirplakJCP2008,MVBerry2009,XChenPRL2010,AdelCampoPRL2013,MMolinerPRL2013}.

In recent years, many theoretical schemes have been proposed on transitionless process \cite{XChenPRA2012,SIbPRL2012,XChenOL2012,SIbPRA2013,ETorronteguiAAMOP2013,SMartPRA2014,MLuPRA2014,LGiannelliPRA2014,AKielyJPB2014,ACSantosSR2015,YLiangPRA2015,XKSongNJP2016,ACSantosPRA2016,XKSongPRA2016,MOkuyamaPRL2016}. For example, Chen and Muga \cite{XChenPRA2012} presented the fast population transfer in three-level systems by invariant-based inverse engineering in 2012.
Ib\'{a}\~{n}ez \emph{et al}. \cite{SIbPRL2012} proposed the multiple Schr\"{o}dinger dynamics (MSDs) method to design alternative and feasible experimental routes for operations in shortcuts to adiabaticity.
In 2013, Ib\'{a}\~{n}ez \emph{et al}. \cite{SIbPRA2013} presented a scheme to improve shortcuts to adiabaticity by iterative interaction pictures.
In 2014, Mart\'{\i}nez-Garaot \emph{et al}. \cite{SMartPRA2014} made a shortcut to adiabaticity in three-level systems via Lie transforms.
Giannelli and Arimondo \cite{LGiannelliPRA2014} presented a work which determines the corrections to the STIRAP pulses required to produce a super-adiabatic transfer in a three-level system.
Kiely and Ruschhaupt \cite{AKielyJPB2014} put forward the population transfer schemes in two- and three-level quantum systems with fast and stable control.
In 2015, Santos and Sarandy \cite{ACSantosSR2015} proposed a general shortcut to controlled adiabatic evolutions through simple time-independent counter-diabatic assistant Hamiltonians.
Liang et al. \cite{YLiangPRA2015} constructed shortcuts to the adiabatic passage for a multiqubit controlled-phase gate.
In 2016, Song \emph{et al}. \cite{XKSongNJP2016,XKSongPRA2016} proposed the shortcuts to adiabatic holonomic quantum computation with TQD algorithm and physically feasible three-level transitionless quantum driving with MSDs.
Due to the simple calculation process and easy to implement in practice, some experimental achievements were demonstrated based on TQD in different systems \cite{MGBason2012,JZhangprl2013,YXDuNC2016,AnNC2016}. For instance, in 2012, Bason \emph{et al}. \cite{MGBason2012} experimentally implemented high-fidelity quantum driving protocols on Bose-Einstein condensates in optical lattices.
In 2013, Zhang \emph{et al}. \cite{JZhangprl2013} experimentally realized the assisted quantum adiabatic passage in the electron spin of a single nitrogen-vacancy center in diamond.
Recently, Du et al. \cite{YXDuNC2016}  demonstrated the stimulated Raman shortcut-to-adiabatic passage with cold atoms in experiment.
An et al. \cite{AnNC2016} experimentally implemented a shortcut to the adiabatic transport of a trapped ion in phase space.

Optomechanical system is a newly-developing solid-state system for studying quantum optics and quantum information processing in last decade \cite{MAspelmeyerRMP2014}. The typical setup for optomechanical systems is composed of a mechanical resonator and an optical or microwave cavity. Many fundamental researches have been studied in optomechanical systems, such as the ground state cooling for a mechanical resonator \cite{FMarquardtPRL2007,IWilsonRaePRL2007,CGenesPRA2008,JDTeufelDonnerNature2011,JChanNature2011}, the observation of strong coupling effects \cite{JDTeufelLiNature2011}, and optomechanically induced transparency \cite{GSAgarwalPRA2010,SWeisScience2010,AHSafaviNaeiniNature2011,CHDongscience}. As a crucial step for quantum information processing and  quantum state engineering, quantum state transfer has attracted much attention in optomechanical systems. Optomechanical systems could make the quantum state transfer between cavities with different wavelengths feasible \cite{LTianPRA2010,KStannigelPRL2010,VFiorePRL2011,ShBarzanjehPRA2011,LTianPRL2012,YDWangPRL2012,ShBarzanjehPRL2012,HKLiPRA2013,ShBarzanjehPRL2015} for the realizable coupling between diverse electromagnetic cavities and a mechanical resonator.
In 2010, Tian and Wang \cite{LTianPRA2010} proposed the scheme to realize optical wavelength conversion of quantum states in the optomechanical system.
At the same year, Stannigel \emph{et al}. \cite{KStannigelPRL2010} presented a new optomechanical transducers for long-distance quantum communication.
In 2011, Fiore \emph{et al}. \cite{VFiorePRL2011} demonstrated experimentally the storage of optical information as a mechanical excitation in a silica optomechanical resonator.
In 2012, Tian \cite{LTianPRL2012} proposed the adiabatic quantum state transfer with high fidelity and pulse transmission scheme in optomechanical systems. Simultaneously, Wang and Clerk \cite{YDWangPRL2012} revisited the problem of using a mechanical resonator to perform the intra-cavity transfer of a quantum state by double swap, adiabatic and itinerant state transfer protocol. The unique advantages of double swap and adiabatic scheme are transitionless and robust to the mechanical dissipation, respectively. The quantum state transfer scheme which can combine the advantages of those two processes is a very meaningful task. Due to optomechanical system can be constructed easily to hybrid structure combined with another system, such as superconducting circuit and spin system. Therefore one can transfer intra-cavity state between different wavelengths and make some applications, such as reversible optical-to-microwave quantum interface \cite{ShBarzanjehPRL2012} and microwave quantum illumination \cite{ShBarzanjehPRL2015}. Therefore the quantum state transfer in optomechanical system has big potential in quantum information processing and quantum computation with unique advantage of scalability.

In this article, we propose the first scheme to achieve the transitionless intra-cavity quantum state transfer based on TQD in optomechanical systems. Our scheme holds the advantages in transitionless, the low dependence of accurately controlling evolution time and being robust to the mechanical dissipation. Our scheme maybe has potential applications in quantum information processing and quantum state engineering, such as reversible frequency conversion, quantum logic gates, and quantum state generation.

This paper is organized as follows:
In Sec.~\ref{sec2}, we introduce the adiabatic quantum state transfer process in optomechanical systems.
In Sec.~\ref{sec3}, we derive the effective M matrix based on TQD algorithm in Heisenberg picture and analyze the result of quantum state transfer in different processes.
In Sec.~\ref{sec4}, we introduce a new physically feasible interaction to realize the TQD quantum state transfer in optomechanical system.
In Sec.~\ref{sec5}, a discussion and a summary are given.

\begin{figure}[!ht]
\begin{center}
\includegraphics[width=8.0cm,angle=0]{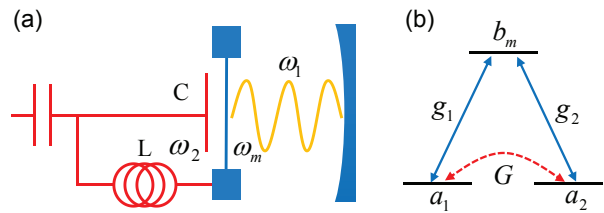}
\caption{(a) Schematic diagram for the optomechanical system. (b) Schematic diagram for the coupling of adiabatic (blue solid-line arrows) and effective coupling Hamiltonian based on TQD algorithm (red dash-line arrow).}\label{fig1}
\end{center}
\end{figure}

\section{adiabatic intra-cavity state transfer in optomechanical systems}\label{sec2}

\begin{table}
  \begin{center}
  \caption{Eigenvalues and corresponding eigenmodes of the matrix $M(t)$.}
    \begin{tabular}{lccccccccccl}
      \hline\hline
      Eigenvalues &&&&&&&&&&& Eigenmodes  \\ \hline
      $\lambda_{1}=0$ &&&&&&&&&&& $\psi_{1}=[-g_{2}/g_{0},0,g_{1}/g_{0}]^{T}$  \\ 
      $\lambda_{2}=-g_{0}$ &&&&&&&&&&& $\psi_{2}=[g_{1}/g_{0},-1,g_{2}/g_{0}]^{T}/\sqrt{2}$  \\ 
      $\lambda_{3}=g_{0}$ &&&&&&&&&&& $\psi_{3}=[g_{1}/g_{0},1,g_{2}/g_{0}]^{T}/\sqrt{2}$ \\ \hline\hline
    \end{tabular}\label{Tab1}
  \end{center}
\end{table}

We consider an optomechanical system shown in Fig.~\ref{fig1}(a), which is composed of two cavity modes coupled to each other via optomechanical forces. After the standard linearization procedure, the Hamiltonian of the system is given by ($\hbar=1$) \cite{CGenesPRA2008,LTianPRL2012,YDWangPRL2012}
\begin{eqnarray}        \label{eq2}
H&\!\!\!=\!\!&\omega_{m}b^{\dag}b+\sum_{i=1,2} \Delta_{i}a^{\dag}_{i}a_{i}+g_{i}(a^{\dag}_{i}+a_{i})(b+b^{\dag}),\;\;\;
\end{eqnarray}
where $a_{i}$$(a^{\dag}_{i})$ $(i=1,2)$ and $b$$(b^{\dag})$ are the annihilation (creation) operator for the $i$-th cavity mode and the mechanical mode, respectively. $\omega_{m}$ is the mechanical frequency. $\Delta_{i}=\omega_{di}-\omega_{i}$ and $g_{i}=g_{0i}\sqrt{n_{i}}$ are the laser detuning and the effective linear coupling strength, respectively. $g_{0i}$ and $n_{i}$ are the single-photon optomechanical coupling rate and intracavity photon number induced by the driving field, respectively.
We consider that both cavity modes are driven near their red sidebands. In the interaction picture, the Hamiltonian becomes (under the rotating-wave approximation) \cite{LTianPRL2012,YDWangPRL2012,HKLiPRA2013}
\begin{eqnarray}      \label{eq2}
H=\sum_{i=1,2}\delta_{i}a_{i}^{\dag}a_{i}+g_{i}(a_{i}^{\dag}b_{m}+b_{m}^{\dag}a_{i}),
\end{eqnarray}
where $\delta_{i}=-\Delta_{i}-\omega_{m}$. Therefore the Heisenberg equation of the system can be derived with
\begin{eqnarray}      \label{eq4}
id\vec{v}(t)/dt=M(t)\vec{v}(t),
\end{eqnarray}
where the vector operator $\vec{v}(t)=[a_{1}(t),b_{m}(t),a_{2}(t)]^{T}$, and the matrix $M(t)$ is expressed as
\begin{eqnarray}      \label{eq4}
M(t)=
\left[
\begin{array}{ccc}
\delta_{1}&g_{1}(t)&0\\
g_{1}(t)&0&g_{2}(t)\\
0&g_{2}(t)&\delta_{2}\\
\end{array}
\right].
\end{eqnarray}
Under the condition that $\delta_{i}=0$, one can get the eigenvalues and eigenmodes of the matrix $M(t)$, shown in TABLE \ref{Tab1}.
The eigenvalue $\lambda_{1}=0$ of the matrix $M(t)$ with eigenmode $\psi_{1}=[-g_{2},0,g_{1}]^{T}/g_{0}$ is a mechanical dark mode that only involves the cavity modes. The adiabatic intracavity quantum state transfer scheme proposed by Tian \cite{LTianPRL2012} is divided into three steps. Shown in Fig.~\ref{fig1}(b), in step 1, one first stores the quantum state in mode $a_{1}$. The two other modes are separable from mode $a_{1}$ in arbitrary single particle states. $g_{1}(0)=0$ and $g_{2}(0)$ start with a large negative value. Therefore, the initial state is $\alpha(0)_{1}=a_{1}(0)$. Step 2, one adiabatically decreases   $g_{2}(t)$ to zero at the end of the process. $g_{1}(t)$ is adiabatically increased from zero to a large positive value. The time of this adiabatical process must satisfy the condition that $T\gg 1/g_{0}$. Step 3, due to the whole process is preserved in the mechanical dark mode, one gets the state with $\alpha(T)_{1}=a_{2}(T)$. According to the Heisenberg equation, $a_{2}(T)=a_{1}(0)$. The result indicates that the initial mode $a_{1}$ has been transferred to mode $a_{2}$ successfully. If one photon in cavity 1, one can transfer it to cavity 2 via an adiabatic process.

\begin{figure}[!ht]
\begin{center}
\includegraphics[width=8.5cm,angle=0]{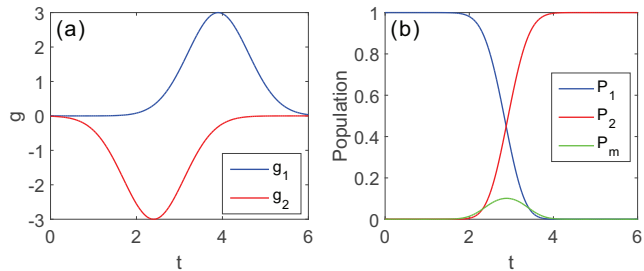}
\caption{Perfect adiabatic quantum state transfer with Gaussian coupling. (a) Diagram for the two Gaussian couplings. (b) The population transfer in the adiabatic process.}\label{fig2}
\end{center}
\end{figure}

Here we choose the Gaussian time-dependence coupling strengths whih are expressed by
\begin{eqnarray}        \label{eqgg}
g_{1}(t)&=&3\ast exp[-(t-2.4+\tau)^{2}],\nonumber\\
g_{2}(t)&=&-3\ast exp[-(t-2.4)^{2}],
\end{eqnarray}
where the coefficients $\tau$ represents the deviation of the time interval between two coupling strengths. As the model could be applied to a wide range of systems, we choose the arbitrary units for the parameters. We fix the coupling strength $g_{2}$ all the time and adjust the  time interval by changing $\tau$ in $g_{1}$. According to our computational result, the time interval designed with $\tau=-0.95$ satisfy the adiabatic condition very well, and one can get a perfect intra-cavity quantum state transfer. We perform the population with respect to time based on adiabatic process in Fig.~\ref{fig2}. The initial state is $|100\rangle$ which indicates that there are just one photon in cavity 1, and no photons in cavity 2, and the mechanical oscillator in the ground state.

\section{intra-cavity state transfer based on TQD algorithm}  \label{sec3}

For universality, firstly we consider a quantum system with an arbitrary time-dependent Hamiltonian $\hat{H}_{0}(t)$. The dynamical process described by the Schr\"{o}dinger equation is given by
\begin{eqnarray}      \label{eq4}
\hat{H}_{0}(t)|n(t)\rangle=E_{n}(t)|n(t)\rangle,
\end{eqnarray}
where $|n(t)\rangle$ and $E_{n}(t)$ are the instantaneous eigenstate and the eigenenergies of $\hat{H}_{0}(t)$, respectively. According to the adiabatic approximation \cite{MBorn1928}, the dynamical evolution of states driven by $\hat{H}_{0}(t)$ could be expressed with
\begin{eqnarray}      \label{eq4}
|\psi_{n}(t)\rangle\!=\!exp\{-i\!\!\int^{t}_{0}\!\!dt'E_{n}(t')\!
-\!\!\!\int^{t}_{0}\!\!dt'\langle n(t')|\dot{n}(t')\rangle\}|n(t)\rangle.\nonumber\\
\end{eqnarray}
Now, we seek a new Hamiltonian $\hat{H}(t)$ based on the reverse engineering approach to satisfy the Schr\"{o}dinger equation
\begin{eqnarray}      \label{eq4}
i|\dot{\psi}_{n}(t)\rangle=\hat{H}(t)|\psi_{n}(t)\rangle.
\end{eqnarray}
Any time-dependent unitary operator $\hat{U}(t)$ is also given by
\begin{eqnarray}      \label{eqU}
i\dot{\hat{U}}(t)=\hat{H}(t)\hat{U}(t),
\end{eqnarray}
and
\begin{eqnarray}      \label{eqH}
\hat{H}(t)=i\dot{\hat{U}}(t)\hat{U}^{\dag}(t).
\end{eqnarray}
In order to guarantee no transition between the eigenstates of $\hat{H}_{0}(t)$ for all time, one should assure that any time-dependent unitary operator $\hat{U}(t)$ has the form
\begin{eqnarray}      \label{eq4}
\hat{U}(t)\!\!&=&\!\!\sum_{n}exp\{-i\!\!\int^{t}_{0}\!\!dt'E_{n}(t')\!
-\!\!\!\int^{t}_{0}\!\!dt'\langle n(t')|\dot{n}(t')\rangle\}\nonumber\\
&&\times|n(t)\rangle\langle n(0)|.
\end{eqnarray}
According to Eq. (\ref{eqH}), the new Hamiltonian is given by
\begin{eqnarray}      \label{eq4}
\hat{H}(t)=\sum_{n}|n\rangle E_{n}\langle n|+i\sum_{n}(|\dot{n}\rangle\langle n|-\langle n|\dot{n}\rangle|n\rangle\langle n|).
\end{eqnarray}
One can find infinitely many Hamiltonian $\hat{H}(t)$ which differ from each other only by phases. We disregard the phase factors \cite{ETorronteguiAAMOP2013,ACSantosPRA2016} and give the simplest Hamiltonian is derived with \cite{MVBerry2009}
\begin{eqnarray}      \label{eq4}
\hat{H}(t)=i\sum_{n}|\dot{n}\rangle\langle n|.
\end{eqnarray}

According to the process introduced before, we calculate the matrix $M(t)$ driving the evolution of cavity and mechanical modes in Heisenberg picture. Analogy to Eq. (\ref{eqU}), we can get the time-dependent unitary operator with
\begin{eqnarray}        \label{eq1}
i\dot{U}(t)=M(t)U(t).
\end{eqnarray}
One can solve the $M(t)$ matrix via the equation given by
\begin{eqnarray}        \label{eqM}
M(t)=i[\frac{\partial}{\partial t}U(t)]U^{\dag}(t)=i\sum_{n}|\frac{\partial n(t)}{\partial t}\rangle\langle n(t)|,
\end{eqnarray}
where $U(t)=\sum|n(t)\rangle\langle n(0)|$. Substitute the all eigenmodes  into Eq. (\ref{eqM}), one can get a new matrix $M(t)$ given by
\begin{eqnarray}      \label{eqM1}
M(t)\!=\!i\!\sum_{n=1}^{3} \dot{\psi_{n}}\psi_{n}^{\dag}
\!=\!i\!
\left[
\begin{array}{ccc}
0&0&-G\\
0&0&0\\
G&0&0\\
\end{array}
\right],
\end{eqnarray}
where $G=(g_{1}\dot{g_{2}}-\dot{g_{1}}g_{2})/g^{2}_{0}$. The result of the matrix $M(t)$ indicates that there should be a direct transition between cavities 1 and 2, as shown in Fig.~\ref{fig1}(b) (the red dash-line transition).

\section{Effective matrix M based on interaction in a largely detuned optomechanical cavity}  \label{sec4}

We present a physically feasible scheme based on largely detuned interaction to realize the transitionless quantum state transfer in same optomechanical system. The new interaction mechanism protects the photon-photon interactions from mechanical dissipations \cite{HKLiPRA2013}. In order to distinguish with the coupling strength of adiabatic process, we rewrite the Hamiltonian with different letters in the interaction picture as
\begin{eqnarray}      \label{eqHn}
H=\sum_{i=1,2}\delta'_{i}a_{i}^{\dag}a_{i}+G_{i}(a_{i}^{\dag}b_{m}+b_{m}^{\dag}a_{i}).
\end{eqnarray}
With the condition $\delta'_{i}\gg G_{i}$, the large energy offsets suppress the transitions between the optical system and the mechanical oscillator \cite{HKLiPRA2013}. Hence, one can adiabatically eliminate the mechanical mode $b$ in Eq. (\ref{eqHn}) and obtain the effective beam-splitter-like Hamiltonian
\begin{eqnarray}        \label{eqbs}
H=\sum_{i=1,2}(\delta'_{i}+\Omega_{i})a_{i}^{\dag}a_{i}
+\Omega(a^{\dag}_{1}a_{2}+a^{\dag}_{2}a_{1}),
\end{eqnarray}
where $\Omega_{i}=G^{2}_{i}/\delta_{i}$ and $\Omega=G_{1}G_{2}(\delta^{-1}_{1}+\delta^{-1}_{2})/2$. We set $\delta'_{1}+\Omega_{1}=\delta'_{2}+\Omega_{2}$ and $\delta'=\delta'_{1}=\delta'_{2}$. In the new interaction picture under the Hamiltonian $H_{0}=\sum_{i=1,2}(\delta'_{i}+\Omega_{i})a_{i}^{\dag}a_{i}$, one can derive the matrix $M(t)$ in the Heisenberg picture with
\begin{eqnarray}      \label{eqM2}
M'(t)\!=\!
\left[
\begin{array}{ccc}
0&0&\frac{G_{1}G_{2}}{\delta'}\\
0&0&0\\
\frac{G_{1}G_{2}}{\delta'}&0&0\\
\end{array}
\right].
\end{eqnarray}
The effective matrix $M'(t)$ shown in Eq. (\ref{eqM2}) is equivalent to the $M$ shown in Eq. (\ref{eqM1}) derived by the TQD algorithm, when
\begin{eqnarray}        \label{eqGG}
\frac{G_{1}G_{2}}{\delta'}=G=\frac{g_{1}\dot{g_{2}}-\dot{g_{1}}g_{2}}{g^{2}_{0}}.
\end{eqnarray}
Hence, we can design the coupling strength $G_{i}$ according to Eq. (\ref{eqGG}) and choose the Gaussian coupling functions in adiabatic process. Therefore, $G_{1}G_{2}=\delta'G'$.

\begin{figure}[ht]
\begin{center}
\includegraphics[width=8.5cm,angle=0]{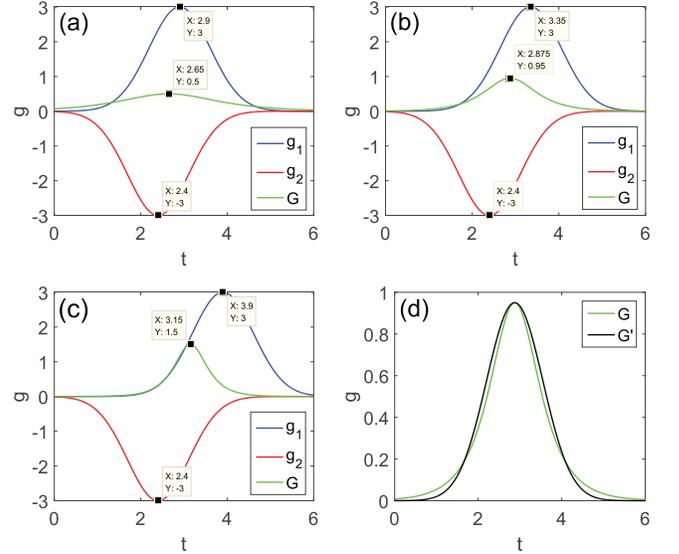}
\caption{Simulation of new Gaussian coupling strength for TQD quantum state transfer. (a) $\tau_{1}=-0.5$. (b) $\tau_{2}=-0.95$. (c) $\tau_{3}=-1.5$. (d) Coupling strength curves for $G$ and $G'$. The parameters are set with $\tau_{2}=-0.95$ and $\alpha=1.1$.}\label{monigaosi}
\end{center}
\end{figure}

The effective coupling strength $G$ is not a typical Gaussian function. In order to simplify the operation in experiment, we choose a new Gaussian coupling strength $G'$ to replace  $G$. The form of $G'=-\tau*exp(-\alpha(t-2.4+\frac{\tau}{2})^{2})$ could be speculated reasonably via incomplete induction shown in Fig.~\ref{monigaosi}. We choose three kinds of time interval $\tau_{1}=-0.5$, $\tau_{2}=-0.95$, and $\tau_{3}=-1.5$ in Fig.~\ref{monigaosi} (a), (b), and (c), respectively. One can find that all the figures satisfy the relationships $-\tau=x_{g2}-x_{g1}=y_{G}$ and $x_{G}(y_{max})=2.4-\frac{\tau}{2}$. Therefore we speculate that the expression of $G'=-\tau*exp(-\alpha(t-2.4+\frac{\tau}{2})^{2})$, where $\alpha$ is used to fit  $G$. We choose $\tau_{2}=-0.95$ and $\alpha=1.1$ and plot the curves of $G$ and $G'$  in Fig.~\ref{monigaosi} (d).

\begin{figure}[!ht]
\begin{center}
\includegraphics[width=8.5cm,angle=0]{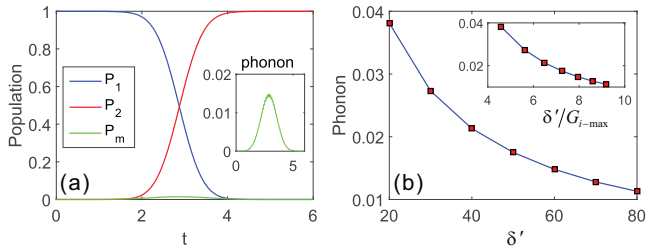}
\caption{(a) Simulation of the transitionless quantum state transfer in largely detuned optomechanical system. The population of phonon is amplified insert. (b) Variation of the maximal average phonon number with detuning. The small figure is variation of the maximal average phonon number relative to proportion between detuning and the maximal coupling strength.}\label{poputqd}
\end{center}
\end{figure}

The other parameters should be chosen to satisfy the large detuning condition $\delta'\gg G_{i}$ and $\delta'_{1}+\Omega_{1}=\delta'_{2}+\Omega_{2}$. Therefore, we set $\delta'=60$. For convenience, we choose $G_{1}=G_{2}=\sqrt{\delta'G'}\approx 7.55\ast exp[-0.55(t-2.875)^{2}]$. Therefore $\delta'/G_{i-max}\approx 8$. With the parameters given before, we plot the evolution of photon and phonon in Fig.~\ref{poputqd}(a) by using Hamiltonian in Eq. (\ref{eqHn}). We find that the Fock state of cavity 1 is transferred perfectly to cavity 2 via the TQD process. The result indicates that the perfect quantum state transfer could be accomplished with the new Gaussian coupling $G'$. The phonon number is suppressed in the whole process and the maximum value is smaller than 0.02 due to the large detuning interaction. Under the large detuning, the population of phonon cannot exist steadily. So after the state of cavity 1 is transferred to phonon, it will be transferred to cavity 2 rapidly. From Fig.~\ref{poputqd}(b), one can find the maximal average phonon number is inversely proportional to the proportion between detuning and the maximal coupling strength. This relationship indicates the reasonability of Eq. (\ref{eqbs}) by adiabatic elimination.

\section{discussion and summary} \label{sec5}

\begin{figure}[!ht]
\begin{center}
\includegraphics[width=8.5cm,angle=0]{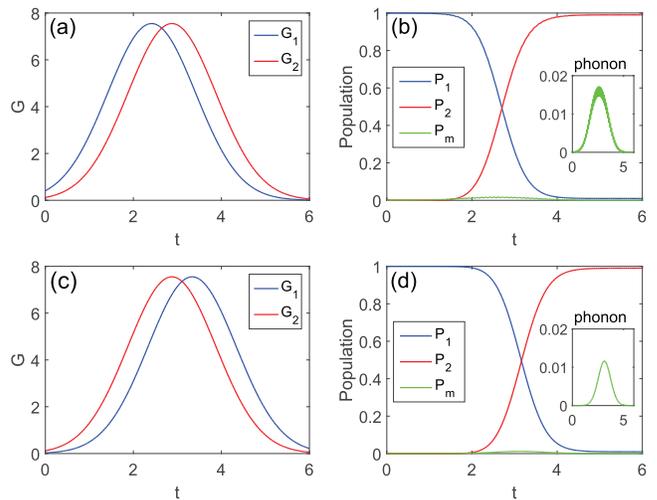}
\caption{The influence on quantum state transfer caused by time interval of two coupling strength. (a) $G_{1}$ begins earlier than $G_{2}$ with $\triangle t=0.46$. (b) The corresponding transitionless quantum state with time interval $\triangle t=0.46$. (c) $G_{2}$ begins earlier than $G_{1}$ with $\triangle t=-0.46$. (d) The corresponding transitionless quantum state transfer with time interval $\triangle t=-0.46$.}\label{bijiao}
\end{center}
\end{figure}

It is hard to accurately control the time interval between two coupling strengths in the actual experiment. Therefore we consider the influence caused by a small deviation of time interval in Fig.~\ref{bijiao}. The $G_{1}$ and $G_{2}$ are designed with no time interval before. We tune the time interval with $\triangle t=\pm0.46$ in Fig.~\ref{bijiao} (a) and (c), respectively. The  Fig.~\ref{bijiao} (b) and (d) are corresponding population evolution. The population of cavity 2 reaches $99\%$ when it keeps stabilize. Fig.~\ref{bijiao} indicates that with a small deviation of the time interval, the quantum state transfer also can be accomplished very well. Our scheme is not very sensitive to the deviation of time interval.

\begin{figure}[!ht]
\begin{center}
\includegraphics[width=8.5cm,angle=0]{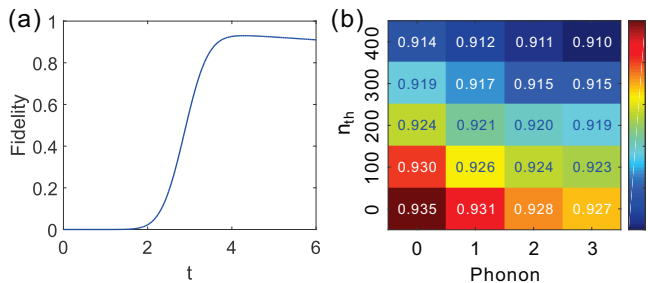}
\caption{(a) Fidelity of TQD process in present of dissipation in optomechanical system. $n_{th}=100$. (b) The maximal fidelity change with initial phonon number and $n_{th}$. Parameters are set with $\kappa_{1}=\kappa_{1}=0.015$, $\gamma_{m}=5\times 10^{-4}$.}\label{fidelity}
\end{center}
\end{figure}

When the dissipation of the mechanical oscillator and the decay of the cavity is taken into consideration. The dynamics of the quantum system described by the Lindblad form master equation is expressed by
\begin{eqnarray}        \label{eq2}
\frac{d\rho}{dt}\!=\!i[\rho,H(t)]\!+\!\kappa_{1}L[a_{1}]\rho\!
+\!\kappa_{2}L[a_{2}]\rho\!+\!\gamma_{m}D[b_{m}]\rho,
\end{eqnarray}
where $\rho$ and $H(t)$ are the density matrix and the Hamiltonian of the optomechanical system, respectively. $\kappa_{1}$ and $\kappa_{2}$ represent the decay rates of cavity 1 and 2, respectively. $\gamma_{m}$ is the mechanical damping rate. $L[A]\rho=(2A\rho A^{\dag}-A^{\dag}A\rho-\rho A^{\dag}A)/2$. $D[A]\rho=(n_{th}+1)(2A\rho A^{\dag}-A^{\dag}A\rho-\rho A^{\dag}A)/2+n_{th}(2A^{\dag}\rho A-AA^{\dag}\rho-\rho AA^{\dag})/2$, where $n_{th}$ is the thermal phonon number of the environment. We choose $\kappa_{1}=\kappa_{1}=0.015$, $\gamma_{m}=5\times 10^{-4}$, and $n_{th}=100$. The final state of the mechanical oscillator do not effect the result which we desire, so we calculate the fidelity with formula $F=\langle 01|tr_{m}[\rho(t)]|01\rangle$. Here, $|01\rangle$ represents the state which there are zero and one photon in cavity 1 and 2, respectively. $tr_{m}[\rho(t)]$ is the reduced density matrix by tracing the mechanical oscillator degree of freedom. The fidelity is plotted in Fig.~\ref{fidelity}(a). The maximal fidelity is $93\%$. We plot the fidelity influenced by the initial phonon number state and $n_{th}$ in Fig.~\ref{fidelity}(b). We change the initial phonon number and $n_{th}$ from $|0\rangle\sim |3\rangle$ and $0\sim 400$, respectively. The maximal fidelity in TQD process just reduces 0.025 (from 0.935 to 0.910). This result indicates that the TQD process can be accomplished with no need the mechanical oscillator in its ground state and it is also robust to the mechanical dissipation.

In the adiabatic quantum state transfer scheme \cite{LTianPRL2012,YDWangPRL2012}, the process based on dark mode which decoupled to mechanical mode is robust to the mechanical dissipation. The advantage of non-adiabatic double swap protocol \cite{YDWangPRL2012} is transitionless in the evolution process. In order to achieve a perfect state result, usually one needs accurately control the terminate time of coupling. In our scheme, it holds both the advantages in the adiabatic and the double swap schemes, and it is not need the high dependence of accurately controlling the terminate time and interval. Robustness to mechanical dissipation derives from the largely detuned interaction mechanism.

In summary, we have proposed a scheme to realize the transitionless quantum state transfer based on the TQD algorithm in optomechanical systems. Also, we have given a physically feasible system with largely detuned interaction. Our scheme can achieve the perfect quantum state transfer with transitionless, the not high dependence of accurately controlling evolution time and sensitiveness to interval between coupling strength, and the robustness to the mechanical dissipation.

\section*{ACKNOWLEDGMENT}

We thank Jing Qiu and Xiao-Dong Bai for useful discussion. This work is supported by
the National Natural Science Foundation of China under
Grants No. 11474026, No. 11475021, and No. 11505007, and
the National Key Basic Research Program of China under Grant No. 2013CB922000.\\

\end{document}